\documentclass[twocolumn,aps,showpacs,superscriptaddress,amssymb,prl]{revtex4}
\usepackage{graphicx}
\usepackage{amsmath}

\newcommand{\ts}{\textstyle}
\newcommand{\mrm}{\mathrm}

\begin{document}

\title{Fluctuations of a weakly interacting Bose-Einstein condensate.}

\author{Zbigniew Idziaszek}
\affiliation{Institute of Theoretical Physics, University of Warsaw,
00-681 Warsaw, Poland}
\affiliation{Center for Theoretical Physics, Polish Academy of Sciences, Aleja Lotnik{\'o}w 32/46, 02-668 Warsaw, Poland}
\author{\L ukasz Zawitkowski}
\affiliation{Center for Theoretical Physics, Polish Academy of Sciences, Aleja Lotnik{\'o}w 32/46, 02-668 Warsaw, Poland}
\author{Mariusz Gajda}
\affiliation{Institute of Physics, Polish Academy of Sciences, Aleja Lotnik{\'o}w 32/46, 02-668 Warsaw, Poland}
\affiliation{Faculty of Mathematics and Sciences, Cardinal Stefan Wyszy\'nski University, Warsaw, Poland}
\author{Kazimierz Rz\c{a}\.{z}ewski}
\affiliation{Center for Theoretical Physics, Polish Academy of Sciences, Aleja Lotnik{\'o}w 32/46, 02-668 Warsaw, Poland}
\affiliation{Faculty of Mathematics and Sciences, Cardinal Stefan Wyszy\'nski University, Warsaw, Poland}

\begin{abstract}
Fluctuations of the number of condensed atoms in a finite-size, weakly interacting Bose gas confined in a box potential are investigated for temperatures up to the critical region. The canonical partition functions are evaluated using a recursive scheme for smaller systems, and a saddle-point approximation for larger samples, that allows to treat realistic size systems containing up to $N \sim 10^5$ particles. We point out the importance of particle-number constrain and interactions between out of condensate atoms for the statistics near the critical region. For sufficiently large systems the crossover from the anomalous to normal scaling of the fluctuations is observed. The excitations are described in a self-consistent way within the Bogoliubov-Popov approximation, and the interactions between thermal atoms are described by means of the Hartree-Fock method.
\end{abstract}


\maketitle

A breakdown of the standard, grand-canonical ensemble to describe fluctuations of an ideal Bose gas and a necessity for
canonical or microcanonical description has been noticed already long time ago \cite{Hauge}, but only in recent decade the problem of fluctuations has received renewed attention due to the experimental achievement of Bose-Einstein condensate (BEC) in ultracold trapped gases. For ideal gases, the canonical and microcanonical
fluctuations have been thoroughly investigated \cite{Politzer,Gajda,MaxDem,Grossmann,Weiss,Grossmann1,HK,Schmidt,Scully}, and several powerful techniques, like the Maxwell Demon ensemble \cite{MaxDem,Weiss,Grossmann1}, have been developed. For interacting particles the fluctuations have been studied mainly within the Bogoliubov approximation \cite{Bogoliubov} of weakly interacting gases \cite{Giorgini,Meier,Kocharovsky,Bhaduri,Xiong,Idziaszek2}, that proved to be extremely successful to describe many other properties of BEC. The exact treatments, so far applied only for one-dimensional systems \cite{Carusotto}, confirmed an excellent agreement with predictions of the Bogoliubov method. We note that some controversy exists about the applicability of the mean-field theory to this problem \cite{Yukalov}, on the other hand, other approaches, like the perturbation theory, lead to qualitatively different results for fluctuations of relatively small condensates \cite{Idziaszek,Illuminati}.
The ultimate verification will be done in experiments. However, to date only the statistics of the total number of atoms has been measured \cite{Raizen}, and a technique involving scattering of short laser pulses has been proposed \cite{Idziaszek3} but not realized

So far the studies of fluctuations in weakly interacting gases have been limited to the regime of low temperatures, and only recently the critical region (close to the critical temperature $T_c$) in a finite-size system has been explored \cite{Svidzynsky}. In this case the Bogoliubov-Popov (B-P) approximation \cite{Popov} has been applied to account for the condensate depletion at finite temperatures and to obtain a description that smoothly interpolates between the degenerate regime below $T_c$ and an ideal gas statistics above $T_c$.

In this Letter we reinvestigate the problem of fluctuations for weakly interacting gas, putting a special emphasis on the interactions of out of condensate atoms, that apart from the critical region, turn out to be important even at moderate temperatures. Following the Bogoliubov-Popov approximation for a uniform Bose gas of $N$ atoms confined in a three-dimensional box of size $L$ with periodic boundary conditions we start with the Hamiltonian:
\begin{equation}\label{Hamiltonian}
\hat{H} = \hat{H}_B + E_{ex}(N,N_0) = \sum_{\textbf{k}\ne 0}\epsilon_{\textbf{k}} \hat{b}_{\textbf{k}}^{\dagger} \hat{b}_{\textbf{k}} + E_{ex}(N,N_0).
\end{equation}
Operators $\hat{b}_{\textbf{k}}=U_{\textbf{k}} \hat{a}_{\textbf{k}} + V_{\textbf{k}} \hat{a}_{-\textbf{k}}^{\dagger}$ are the Bogoliubov quasiparticle annihilation operators, obeying Bose commutation relations $[\hat{b}_{\textbf{k}},\hat{b}_{\textbf{k}^{\prime}}^{\dagger}]=\delta_{{\textbf{k}},{\textbf{k}}^{\prime}}$,
$\hat{a}_{\textbf{k}}$ represent annihilation operators for a mode with quantized momentum $\hbar {\textbf{k}}$. The celebrated Bogoliubov-Popov energy spectrum:
\begin{equation}\label{energy spectrum}
\epsilon_{\textbf{k}}= \sqrt{(\epsilon_{\textbf{k}}^0+ g n_0)^2-(g n_0)^2}
\end{equation}
depends on the condensate density $n_0 =N_0/V$. Here, $\epsilon_{\textbf{k}}^0 = 4\pi^2\hbar^2 k^2/mL^2$ is the kinetic energy of a mode $\textbf{k}$, $m$ is the mass of atoms, $g=4 \pi \hbar^2 a/m$ is the interaction strength, and $a$ is the $s$-wave scattering length characterizing the contact potential  $V(\textbf{r}-\textbf{r}^\prime)=g\delta^{(3)}(\textbf{r}-\textbf{r}^\prime)$.
Bogoliubov coefficients satisfy equations:
$U_{\textbf{k}}^2+V_{\textbf{k}}^2=(g n_0 + \epsilon_{\textbf{k}}^0)/\epsilon_{\textbf{k}} \equiv W_{\textbf{k}}$ and $U_{\textbf{k}}^2-V_{\textbf{k}}^2=1$ \cite{Popov}. Finally, $E_{ex}(N,N_0)$ describes the interaction energy between out of condensate atoms, which in the B-P model can be calculated on the level of Hartree-Fock (HF) approximation: $E_{ex}(N,N_0) = \frac{g}{2V} N_{ex}^2$ \cite{Pethick}, with $N_{ex} = N - N_0$ denoting the number of atoms in excited ($\textbf{k} \neq 0$) modes. This corresponds to taking only secular part of interactions between thermal atoms. The considered Hamiltonian neglects a finite life-time of quasiparticle excitations arising from interaction between quasiparticles \cite{Beliaev}.

The canonical-ensemble partition function for a system with $N$ atoms and temperature $k_B T=1/\beta$ is
\begin{equation}\label{partition}
Z(N,\beta)=\sum_{N_{ex}=0}^{N}\sum_{n_1=0}^{\infty}\ldots \sum_{n_\textbf{k}=0}^{\infty}\ldots
e^{-\beta E} \delta_{N_{ex},\overline{N}_{ex}},
\end{equation}
where $n_{\textbf{k}}$ are populations of quasiparticle excitations,
$E=\sum_{\textbf{k}\ne 0}\epsilon_{\textbf{k}}(N_0)\ n_{\textbf{k}} + E_{ex}(N_{ex})$ is the energy and $\overline{N}_{ex}=\sum_{\textbf{k}\ne 0} n_{\textbf{k}} W_{\textbf{k}} +V_{\textbf{k}}^2$ is the number of thermal atoms of a given configuration of excitations, that differs from the total number of excitations due to the Bogoliubov transformation \footnote{Atomic populations corresponding to quasiparticle excitations are not integers and we have to apply some binning scheme (see e.g. \cite{Idziaszek2})}. Although the condensate mode does not appear in the sum in the Hamiltonian, its population affects the energy spectrum and the interaction energy of atoms in excited modes. In order to enforce the constrain on the
total number of particles rigorously, we keep the energy spectrum dependent on the actual number of condensed atoms, as follows from Eq.~\eqref{partition}.
We calculate the conditional statistical partition function:
\begin{equation}\label{Z_N0^Nex}
Z_{N_0}(N_{ex})=\sum_{n_1=0}^{\infty}\ldots \sum_{n_\textbf{k}=0}^{\infty}\ldots
e^{-\beta E} \delta_{N_{ex},\overline{N}_{ex}},
\end{equation}
which corresponds to a case with $N_0$ condensed atoms and $N_{ex}$ thermal atoms.
In terms of these functions the probability of finding $N_0$ condensed atoms is $P(N_0)=\frac{Z_{N_0}(N-N_0)}{Z}$ and $Z=\sum_{N_0=0}^N Z_{N_0}(N-N_0)$.

The recurrence algorithm used in our calculations is an enhanced version of the earlier algorithm applied to the ideal Bose gas (IBG) \cite{Weiss}, and it will be presented in details elsewhere. It makes use of the fact that $Z_{N_0}(N_{ex})$ treats the number of condensed and thermal atoms as independent variables and the number of condensed atoms becomes a parameter. As an intermediate step one obtains the following result for the mean number of quasiparticle excitations in mode $\textbf{q}$ provided that there are $N_0$ condensed atoms and $N_{ex}$ thermal atoms in the system
\begin{equation}\label{<nq>}
\begin{split}
\langle n_{\textbf{q}}\rangle _{N_0}^{N_{ex}}=\sum_{l=1}^{\infty} e^{-\beta l \epsilon_{\textbf{q}}}\frac{Z_{N_0}\left(N_{ex}-lW_{\textbf{k}}\right)}{Z_{N_0}\left(N_{ex}\right)}.
\end{split}
\end{equation}

We calculate the mean condensate population $\langle \hat{N}_0 \rangle=\sum_{N_0=0}^N P(N_0) N_0$
and its fluctuations $\langle \delta^2 \hat{N}_0 \rangle=\langle \hat{N}_0^2 \rangle - \langle \hat{N}_0 \rangle^2$. In the canonical ensemble $\langle \hat{N}_0 \rangle= N - \langle \hat{N}_{ex} \rangle$ and $\langle \delta^2 \hat{N}_{0} \rangle = \langle \delta^2 \hat{N}_{ex} \rangle$. The fluctuations, can be written as a sum of two contributions: $\langle \delta^2 \hat{N}_{ex} \rangle = \langle \delta^2 \hat{N}_{ex} \rangle_{T} + \langle \delta^2 \hat{N}_{ex} \rangle_{Q}$. The first term represents thermal fluctuations, that we calculate from the probability distribution $P(N_0)$
\begin{align}\label{fluctuations1}
\langle \hat{N}_{ex}^2 \rangle_{T} & =
\sum_{\textbf{k,q}\ne 0} W_\textbf{k} W_\textbf{q} \left(\langle \hat{n}_{\textbf{k}}\hat{n}_{\textbf{q}}\rangle - \langle\hat{n}_{\textbf{k}}\rangle \langle\hat{n}_{\textbf{q}}\rangle\right), \\
& = \sum_{N_0=0}^{N}\!\!N_0^2 P(N_0) - \langle N_0 \rangle.
\end{align}
The second term, $\langle \delta^2 \hat{N}_{ex} \rangle_{Q}$ describes the quantum part of the fluctuations, a non vanishing component at $T=0$ in an interacting gas \cite{Giorgini}.
They result from the Bogoliubov transformation applied to the quantum average of the $\hat{N}_{ex}^2$ operator:
\begin{equation}\label{fluctuations2}
\langle \delta \hat{N}_{ex}^2 \rangle_{Q} = 4 \sum_{\textbf{k}\ne 0} \left\langle U_{\textbf{k}}^2 V_{\textbf{k}}^2 (\hat{n}_{\textbf{k}}\hat{n}_{-\textbf{k}}\!+\hat{n}_{\textbf{k}}\!+\!\ts{\frac12})\right\rangle,
\end{equation}
In the equations above, the average of an arbitrary operator can be expressed in terms of conditional averages: $\langle \hat{X}\rangle =\sum_{N_0=0}^{N} \langle \hat{X}\rangle _{N_0}^{N-N_0} P(N_0)$, with the mean occupation numbers $\langle n_{\textbf{q}}\rangle _{N_0}^{N_{ex}}$ given by \eqref{<nq>}, and the correlation of modes with the opposite momenta
$\langle \hat{n}_{\textbf{k}}\hat{n}_{-\textbf{k}}\rangle _{N_0}^{N_{ex}}=
\sum_{l,j=1}^{\infty} e^{-\beta (l+j) \epsilon_{\textbf{q}}} Z_{N_0}\left(N_{ex}-(l+j)W_{\textbf{q}}\right)/Z_{N_0}\left(N_{ex}\right)$.

From the practical point of view the recursive method is applicable for systems of maximum a few hundred particles. For larger $N$, the calculations become numerically very demanding, and to treat larger samples we have developed a semi-analytical approach, that is based on saddle-point approximation to the contour-integral representation of $Z(N_{ex})$, known in the literature as Darwin-Fowler method \cite{Huang}. Derivation proceeds basically in the same manner as for an ideal gas \cite{MaxDem,HK} and yields
\begin{equation}
\label{Zappr}
Z_{N_0}(N_{ex},\beta) \approx
\frac{\displaystyle \Xi_{N_0}(z_0,\beta)}{\displaystyle z_0^{N_{ex}} \sqrt{2 \pi \frac{\partial^2}{\partial \lambda_0^2} \ln \Xi_{N_0}(z_0,\beta)}},
\end{equation}
where $\Xi_{N_0}(z,\beta)$ is the grand-canonical partition function for the excited subsystem,
$z_0 = e^{\lambda_0}$ denotes the position of the saddle point, determined by
$\langle \hat{N}_{ex} \rangle_\mrm{GC}  = N_{ex}$,
with $\langle \hat{N}_{ex} \rangle_\mrm{GC} \equiv - \partial_{\lambda_0} \ln \Xi_{N_0}(z_0,\beta)$ denoting the grand-canonical expectation value for the number of excited atoms.
In analogy to the ideal gas, $\Xi_{N_0}(z,\beta)$ can be written in a closed form
\begin{equation}
\Xi_{N_0}(z,\beta) = \prod_{{\bf k} \neq 0} z^{V_{\bf k}^2}\left[1-z^{W_{\textbf{k}}} \exp( -\beta \epsilon_{\bf k})\right]^{-1}.
\end{equation}
A similar saddle-point method can be applied to determine $\langle \hat{n}_{\textbf{k}}\rangle$ and $\langle \hat{n}_{\textbf{k}}\hat{n}_{-\textbf{k}} \rangle$ entering formula \eqref{fluctuations2} for $\langle \delta^2 \hat{N}_0^2 \rangle$.

\begin{figure}[t]
\includegraphics[width=0.48\textwidth]{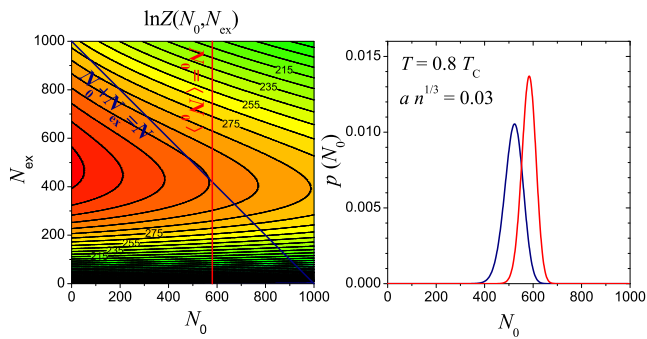}
\caption{\small{
(Colors on-line) Logarithm of the canonical partition function as a function of $N_0$ and $N_{ex}$ (a), along with cross-sections yielding the probability distributions of the condensate (b) for rigorous (blue) and average spectrum (red). Parameters are:
$N=1000$, $an^{1/3}=0.03$ and $T=0.8 T_c$.}}
\label{Fig:Contour}
\end{figure}

This way we have obtained a scheme that allows us to calculate statistical properties of the weakly-interacting condensate at all temperatures. While we keep only the HF contribution to interactions between quasiparticles we otherwise preserve the number of atoms throughout the calculations, which requires inclusion of the energy spectrum dependent on the actual number of condensed atoms. This can be contrasted to the common approximation assuming the energy spectrum dependent on the mean number of condensed atoms: $\epsilon_{\textbf{k}}(N_0)=\epsilon_{\textbf{k}}(\langle N_0\rangle)$ \cite{Kocharovsky,Svidzynsky,PomeauRica_nonlinearwaves}, used in a simple but useful models of thermal equilibrium with Bose-populated excitations \cite{Giorgini,Idziaszek}. This way one would obtain different formulas for $Z=\sum_{N_0=0}^N Z_{\langle N_0\rangle}(N-N_0)$ and $P(N_0)=\frac{Z_{\langle N_0\rangle}(N-N_0)}{Z}$, which, being numerically less demanding, require self-consistent determination of $\langle N_0\rangle$. However, such a seemingly natural simplification leads to observable distortions of the results. This is illustrated in Fig.~\ref{Fig:Contour} presenting the canonical partition function of the excited subsystem in the parameter space $(N_0,N_{ex})$. Inclusion of the excitation spectrum dependent on the actual number of condensed atoms, correspond to performing a cut along a line $N_0 + N_{ex} = N$, whereas the approximation assuming average spectrum corresponds to a cut along  $N_0 = \langle N_0 \rangle = const$, with $\langle N_0 \rangle$ determined in a self consistent way. These two approaches yield probability distributions of the number of condensed atoms (see right panel) that differ both in the position of the maximum and the width of the peak, that determine the values of $\langle N_0 \rangle$ and $\langle \delta^2 \hat{N}_{0} \rangle$, respectively.

\begin{figure}[t]
\includegraphics[width=0.45\textwidth]{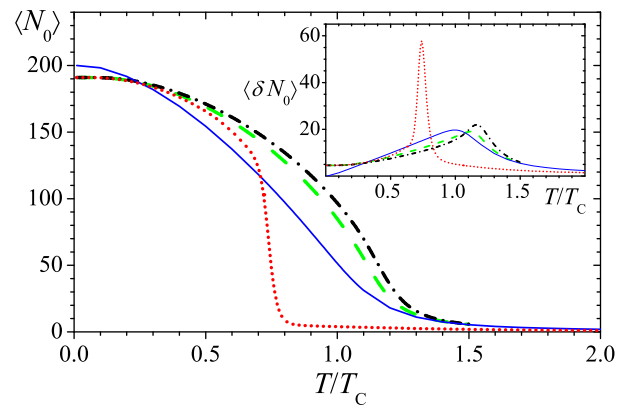}
\caption{\small{
(Colors on-line) Condensate population and fluctuations (inset) versus scaled temperature $T/T_c$ for a system of $N=200$, and $an^{1/3}=0.1$, described by the Bogoliubov-Popov Hamiltonian $H_B$ with (\textit{black dot-dash}) and without (\textit{red dots}) inclusion of interactions between of out of condensate atoms $E_{ex}$. Blue solid line represents a corresponding ideal gas, while green dash line shows the results for the model assuming average spectrum $\varepsilon_\textbf{k}(\langle N_0 \rangle)$ without $E_{ex}$ term.}}
\label{Fig:N200}
\end{figure}

The comparison of the mean condensate populations and fluctuations, for relatively small system
of $N=200$, and $an^{1/3}=0.1$, calculated for rigorous and average spectrum is presented in Fig~\ref{Fig:N200}. The results obtained in the model with rigorous spectrum and the HF contribution to a thermal atoms interaction differ substantially from the mean occupation and fluctuations obtained when the thermal atoms interaction is totally neglected.
On the other hand we note that the two other approaches for an interacting gas lead to rather similar results, and some discrepancies can be only observed in behavior of fluctuations close to the critical temperature.

\begin{figure}[t]
\includegraphics[width=0.48\textwidth]{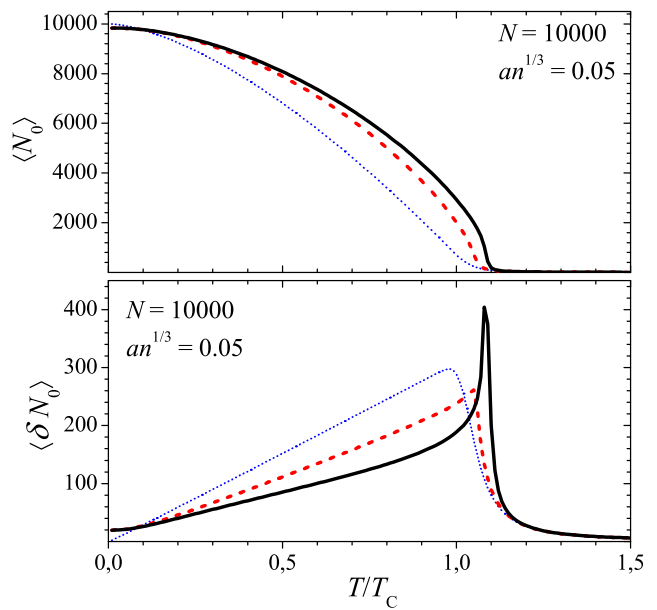}
\caption{\small{
(Colors on-line) Condensate population (upper panel) and fluctuations (lower panel) versus scaled temperature $T/T_c$ for a system with $N=10000$, $an^{1/3}=0.05$ obtained for:
(\textit{black solid}) $H_B+E_{ex}$ with $N_0$-dependent spectrum,
(\textit{red dash}) $H_B$ with average spectrum approximation,
(\textit{blue thin solid}) an ideal gas.}}
\label{Fig:N10000}
\end{figure}

The situation changes, however, for larger systems (see Fig~\ref{Fig:N10000}).
For sufficiently large system of $N=10000$ atoms, and $an^{1/3}=0.05$ inclusion of the rigorous spectrum together with $E_{ex}$ significantly affect the condensate statistics. This is more evident in the case of fluctuations, that in our model remain much smaller than the fluctuations calculated in the model assuming average spectrum, even at temperatures much smaller than the critical one.

\begin{figure}[t]
\includegraphics[width=0.48\textwidth]{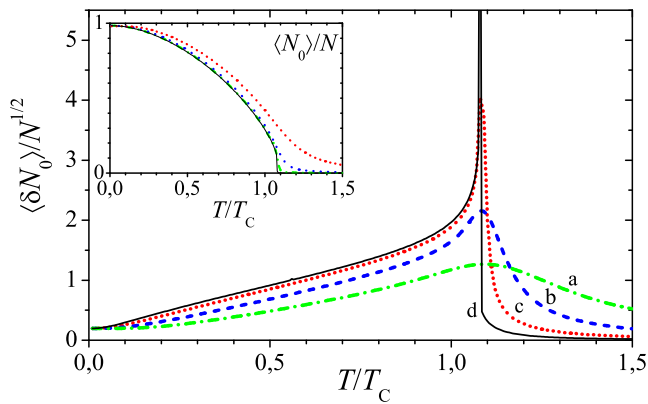}
\caption{\label{Fig:CompN}\small{
(Colors on-line) Normalized condensate fluctuations $\langle \delta \hat{N}_0\rangle/N^{1/2}$ and condensate fraction (inset)
versus scaled temperature $T/T_c$ for fixed $an^{1/3}=0.05$ but varying $N$:
(a, \textit{red dots}) $N=100$,
(b, \textit{blue dash}) $N=1000$,
(c, \textit{green dot-dash}) $N=10^4$,
(d, \textit{black solid}) $N=10^5$.}}
\end{figure}

Finally we have verified how the mean condensate population and its fluctuations depend on the size of the system, while keeping the interaction parameter $a n^{1/3}$ fixed. Fig.~\ref{Fig:CompN} presents the results for the rigorous spectrum including the excited atom interactions, for
$an^{1/3}=0.05$ and the number of particles varying from $N=100$ to $N=10^5$. While the size of the system increases, fluctuations tend to be proportional to $\sqrt{N}$, so they become normal. On the other hand, for small systems the scaling remains anomalous. This result shows that anomalous scaling ($\delta N_0 \sim N^{2/3} $), predicted within the Bogoliubov method neglecting the interactions of thermal atoms \cite{Giorgini,Kocharovsky,Xiong,Idziaszek2}, holds only for relatively small number of atoms.

One observes that for large systems the fluctuations exhibit a high and narrow peak close to $T_c$. We are not sure of its physical significance since the B-P spectrum is questionable so close to the critical temperature. We point out that position of this peak, that define some characteristic temperature for our model, remain fixed in the thermodynamic limit ($a n^{1/3} = const$, $N \rightarrow \infty$). We have verified that the position of the peak vary with interactions as $\Delta T_{ch}/T_c = 1.56 an^{1/3}$ , which is similar to the shift of the critical temperature in an interacting gas: $\Delta T_{c}/T_c = 1.29 an^{1/3}$ \cite{Kashurnikov}.

In this Letter we have presented the most complete to date discussion
of the statistical properties of BEC confined to a box. In particular
we have pointed out the importance of the strict enforcement of the
particle number conservation and of the interactions between thermal atoms. These two new elements, that have been neglected in the previous approaches, turn out to be relevant not only close to the critical region, but also at moderate temperatures, affecting the scaling properties of fluctuations in sufficiently large systems.

\acknowledgments
Z.I., M.G. and K.R. acknowledge support of the Polish Government
Research Grant for 2006-2009, L.Z. acknowledges support of the Polish Government Research Grant for 2006-2008 (No. N202 178 31/3918).


\begin{thebibliography}{99}


\bibitem{Hauge} E.H. Hauge, Physica Nor. {\bf 4}, 19 (1969); R.M. Ziff, G.E. Uhlenbeck, and M. Kac, Phys. Rep.
{\bf 32}, 169 (1977); E. Schr{\"o}dinger, {\it Statistical Thermodynamics} (Dover Publ., New York, 1989).

\bibitem{Politzer} H.D. Politzer, Phys. Rev. A {\bf 54}, 5048 (1996).

\bibitem{Gajda} M. Gajda and K. Rz\c{a}\.{z}ewski, Phys. Rev. Lett. {\bf 78}, 2686 (1997).

\bibitem{MaxDem} Navez {\it et. al.}, Phys. Rev. Lett. {\bf 79}, 1789 (1997).

\bibitem{Grossmann} S. Grossmann and M. Holthaus, Phys. Rev. Lett. {\bf 79}, 3557 (1997).

\bibitem{Weiss} C. Weiss and M. Wilkens, Opt. Ex. {\bf 1}, 272 (1997).

\bibitem{Grossmann1} S. Grossmann and M. Holthaus, Opt. Ex. {\bf 1}, 262 (1997).

\bibitem{HK} M. Holthaus, E. Kalinovski, Ann. Phys. {\bf 276}, 321 (1999).

\bibitem{Schmidt} H.-J. Schmidt and J. Schnack, Physica A {\bf 260}, 479 (1998).

\bibitem{Scully} M.O. Scully, Phys. Rev. Lett. {\bf 82}, 3927 (1999).

\bibitem{Bogoliubov} N. Bogoliubov, J. Phys. USRR {\bf 11}, 23 (1947).

\bibitem{Giorgini} S. Giorgini, L. P. Pitaevskii, and S. Stringari, Phys. Rev. Lett. {\bf 80}, 5040 (1998).

\bibitem {Meier} F. Meier, and W. Zwerger, Phys. Rev. A {\bf 60}, 5133 (1999).

\bibitem{Kocharovsky} Kocharovsky {\it et. al.}, Phys. Rev. Lett. {\bf 84}, 2306 (2000).

\bibitem{Bhaduri} Bhaduri {\it et. al.}, J. Phys. B {\bf 35}, 2817 (2002).

\bibitem {Xiong} Xiong {\it et. al.}, Phys. Rev. A {\bf 65}, 033609 (2002).

\bibitem{Idziaszek2} Z. Idziaszek, Phys. Rev. A {\bf 71}, 053604 (2005).

\bibitem{Carusotto} I. Carusotto, and Y. Castin, Phys. Rev. Lett. {\bf 90}, 030401 (2003).

\bibitem{Yukalov} V.I. Yukalov, Phys. Lett. A {\bf 340}, 269 (2005).

\bibitem{Idziaszek} Z. Idziaszek {\it et. al.}, Phys. Rev. Lett. {\bf 82}, 4376 (1999).

\bibitem{Illuminati} F. Illuminati, P. Navez, and M. Wilkens, J. Phys. B {\bf 32}, L461 (1999).

\bibitem{Raizen} C.S. Chuu {\it et al.}, Phys. Rev. Lett. {\bf 95}, 260403 (2005).

\bibitem{Idziaszek3} Z. Idziaszek, K. Rz\c{a}\.{z}ewski, and M. Lewenstein, Phys. Rev. A {\bf 61}, 053608 (2000).

\bibitem{Svidzynsky} A.A. Svidzinsky, and M.O. Scully, Phys. Rev. Lett. {\bf 97}, 190402 (2006).

\bibitem{Popov} V.N. Popov, {\it Functional Integrals in Quantum Field Theory and Statistical Physics} (Reidel Publishing Company, Dordrecht, 1983), chapter 6.

\bibitem{Pethick} C. J. Pethick, H. Smith, {\it Bose-Einstein Condensation in Dilute Gases}, (Cambridge Univ. Press, 2000).

\bibitem{Beliaev} S. T. Beliaev, Zh. Eksp. Teor. Fiz. {\bf 34}, 433 (1958) [Sov. Phys.-JETP {\bf 34 (7)}, 299 (1958)].



\bibitem{Huang} K.\ Huang, {\it Statistical Mechanics} (John Wiley and Sons, New York, 1987), chapter 9.

\bibitem{NoteBEC} For IBG see e.g. \cite{MaxDem,HK}.



\bibitem{PomeauRica_nonlinearwaves} C. Connaughton \textit{et al.}, Phys. Rev. Lett. {\bf 95}, 263901 (2005).

\bibitem{Kashurnikov} V. A. Kashurnikov, N. V. Prokof'ev, and B. V. Svistunov, Phys. Rev. Lett. {\bf 87}, 120402 (2001).


\end{thebibliography}
\end{document}